\documentclass[aps,a4paper,pre,floatfix,showkeys,twocolumn]{revtex4}
\usepackage{graphicx}
\usepackage{amsmath}
\usepackage{amssymb}
\usepackage{wasysym}
\usepackage{fdsymbol}
\usepackage{setspace}
\usepackage{color}
\usepackage[dvipsnames]{xcolor}
\usepackage{multirow}

\begin{document}

\preprint{AIP/123-QED}

\title{Influence of homeostatic mechanisms of bacterial growth and division on structural properties of microcolonies. A computer simulation study.}

\author{Andr\'es Delgado-Campos}
\affiliation{Department of Physical, Chemical and Natural Systems, Pablo de Olavide University, 41013 Seville, Spain}%

\author{Alejandro Cuetos}

 \email{acuemen@upo.es}
 \affiliation{Department of Physical, Chemical and Natural Systems, Pablo de Olavide University, 41013 Seville, Spain}%

\date{\today}

\begin{abstract}
Bacterial growth and division generally occur by the process known as binary fission, in which the cells grow polarly until they divide into two daughter cells. Although this process is affected by factors that introduce stochastic variability in both growth rate and daughter cell length, the fact is that the size distribution in growing bacteria remains stable over time. This suggests the existence of homeostatic mechanisms that contribute to maintaining a stable size distribution. Those known as sizer and adder stand out among these mechanisms whose relevance is not entirely determined. In this work, computer simulations using an agent-based model, are used to study the effect of these homeostatic mechanisms on the geometrical and structural properties of the developing microcolonies, focusing on the early stages of its development. Also, it was examined the effect of linear or exponential dependence with the time of cellular growth on these properties. From our study, we deduce that these mechanisms do not have a noticeable impact on the properties studied, which could be due to the importance that stochastic factors play in the cell division and growth process. In addition, we discuss how competition between cell growth and diffusion is a key aspect in explaining the structure and geometry of developing bacterial microcolonies. The results of the study will help to clarify which processes and parameters should be considered relevant when designing simulation models.
\end{abstract}


\maketitle


\section{INTRODUCTION}

Cell size is one of the key aspects of bacterial life. It is essential to understand properties such as the surface-volume relationship, cytoplasmic changes, or reproduction processes \cite{TRA11,CHI12,LEV15}. An important feature is that the size changes throughout the bacterial life cycle. Thus, most bacteria rely on binary fission for reproduction, a process in which the individual cell grows from the original length to a certain threshold then the bacterium divides into two daughter bacteria \cite{MAR14}. How this lengthening process is executed, and when and at what size the bacteria carry out this division remains a matter of debate. For instance, is well established since the 1960s that bacteria grow exponentially from their initial size \cite{SCH58,WAN10,GOD10,IYE14,YUF17}, although other authors have reported cases with linear growth  \cite{KUB68,BOU21}, or situations with biphasic growth dynamics \cite{SAL20}. Reshes and colleagues have suggested complex cell shape dynamics, with the possibility of bilinear or trilinear growth regimes, strongly conditioned by the process of septum formation \cite{RES08}.

Another issue on which there is less consensus, with numerous recent publications on the subject, is the mechanism that controls the cell size. That is, considering the stochasticity of the reproductive process in bacteria, how the size distribution of newborn individuals or of the entire cell set is kept stable \cite{TAH15}. The truth is that under steady-state conditions, populations of bacteria tend to maintain stable cell size distribution, with parameters within a narrow range. A variety of homeostatic mechanisms have been proposed to take part in this size control. These processes could be grouped into three limit cases: \textit{sizer}, \textit{timer} and \textit{adder} mechanisms \cite{FAC17,SAU16,TAH15}. In the sizer mechanism, the cells divide when they reach a given size, regardless of the initial size. In the adder mechanism all the cells add the same size increase, again independent of the size at birth. In timer, bacteria grow for a fixed time duration. 
			
If there are no stochastic factors in the growth and division process, for instance dispersion in growth rate, division size, total increment, or time duration, all cells would grow synchronously, with universal initial and division lengths. In this case, the size distribution would remain constant throughout successive generations.  Alternatively, if a cell exhibits a divergence in the initial or final elongation due to stochastic effects or environmental conditions, their descendants will correct the divergence in one generation if the sizer mechanism is applied, while for adder behavior it would take multiple generations to correct the divergence. Timer mechanism is less efficient, showing a weak homeostatic response \cite{FAC17}. As mentioned, there is no consensus on which is the dominant mechanism in eukaryotes or prokaryotic, or if any of them is universal. In any case, the timer mechanism is the one that is considered the least relevant \cite{FAC17,FAC19}, while some authors have suggested the possibility of hybrid mechanisms \cite{FAC19,XIA20}. 

The aim of this article is to provide information on how the different mechanisms outlined above affect the shape and structural properties of growing microcolonies. We have focused on the adder and sizer mechanisms, studying the effect of linear or exponential growth of each individual bacterium. Our goal is not to provide information to support the validity of the various mechanism but to analyze how they can influence the collective properties of early biofilms, termed microcolonies. This study is primarily motivated by two aspects. The detection of specific characteristics in the microcolony's structural properties, if caused by sizer and/or adder behavior, could help determine which homeostatic mechanisms is at work. Furthermore, when developing theoretical and simulation models to investigate the characteristics of bacterial populations, it is essential to know the magnitude of the different factors that affect cell reproduction. We aim to shed light on the importance of choosing the appropriate scenario in theoretical and simulation studies.

Despite some recent and relevant publications that have attempted to explain the dominant homeostatic mechanism, this remains an open question due to the difficulty of getting conclusive experimental results. This is why computer simulation could be a very useful tool in this regard. At the molecular and colloidal level, computer simulation techniques have become well established as tools for routine use in physicochemistry and materials science.  Furthermore, during the last decades, many computational studies have been carried out to investigate the development of bacterial biofilms \cite{WAN10,DEL18,ACE18}, tumors \cite{KAT11,SAV12,MET19}, or tissues \cite{JON12,LIE15,CAM17,LOB21}. A strategy very commonly employed in these studies is defined as individual-based models (IbM). In this approximation, it is proposed that the characteristics of the growing process of a cell community (from microcolonies to biofilms for bacteria) can be described by considering the main features of each bacterium and how they interact with each other \cite{KRE01,WAN10,LIE15}. These models are very similar to molecular dynamical simulation approaches, to the point that it might be appropriate to classify them as cell simulation approaches. A key difference between molecular simulation and cell simulation is that, in the latter case, the growth and division of individual cells play an important role. 

Based on this approach, we recently developed a model for the study of the early stages of the development of bacterial biofilms \cite{ACE18}. In these early stages, microcolonies can be considered as two-dimensional structures. The number of cells reached in our simulations (around 300) may be sufficiently high that this approximation cannot be considered realistic. In any case, we think that for the study of the influence of different homeostatic mechanisms on microcolonies the results will remain valid, and we do not consider the processes of evolution to three-dimensional structures. In addition, there is an extensive literature on the development of microcolonies confined between two planes to maintain the two-dimensional character \cite{YOU18,VOLF08}. Using our model, which explicitly includes the rod shape of bacteria and the growth and division of individual cells, we calculated some structural characteristics of microcolonies, focusing on how they are affected by the competition between cell growth and cell diffusion. In subsequent studies, we extended our model by explicitly introducing the presence of non-adsorbing polymers \cite{LOB21}, or, utilizing the same basic assumptions, to analyze the development of tissues such as the fly-eye \cite{LOB21b}. 

In this study, we have extended our previous IbM model \cite{ACE18} to study the effects of linear vs exponential elongation and sizer or adder homeostatic mechanisms on the structural properties of early bacterial microcolonies when they are considered two-dimensional. As we will see, the different scenarios do not show significant differences in the structural quantities calculated. This may be caused by stochastic dispersion introduced in some individual bacterial characteristics, such as the growth speed, or elongation at the division.

This article is arranged as follows. In Section \ref{methods} we describe the model and simulation methodology employed. Then in Section \ref{results} we present and discuss the results about the influence of each growing scenario in the structure of the microcolony. Finally, we present our conclusions.

\section{METHODS}\label{methods}

In order to explore the effects of the different bacterial reproduction mechanisms, we have used a very similar methodology to that described in \cite{ACE18}. Thus, the first stages of microcolonies growth, when it can be considered two-dimensional, were modeled using an Individual Based Model (IbM) \cite{KRE01,WAN10,LIE15}. In our model, we have assumed that the bacteria lack the capability of active motion, being displaced only by the effect of the interaction with other bacteria as well as through passive diffusion. More specifically, a rod-like bacteria is modelled as a bidimensional spherocylinder. This shape consists of a cylinder of instantaneous elongation $L$ capped by two hemispheres of diameter $\sigma$. During the simulation, the elongation of the cylinder will change over time, while the diameter is going to remain constant throughout the evolution of the system and for all bacteria. Accordingly, the instantaneous aspect ratio of the cell is $L^*=L/\sigma +1$. As in \cite{ACE18}, we have considered that bacteria interact with each other via the soft spherocylindrical potential \cite{EAR01,CUE17}:

\begin{equation}\label{eq1}
U_{ij}\,= \left\{ \begin{array}{cc}
4 \epsilon_{ij} \left[ \left(\frac{1}{d^*_m}\right)^{12} -
\left( \frac{1}{d^*_m}\right)^{6} + \frac{1}{4} \right] & ~~   d^*_{m} \leq \sqrt[6]{2}
\\
0 & ~~ d^*_{m} > \sqrt[6]{2} \end{array} \right.
\end{equation}

where $i$ and $j$ are generic particles (bacteria). $d_m^*=d_m/\sigma$ is the minimum distance between them \cite{VEG94}. With this interaction potential, we pretend to mimic the steric repulsion between bacteria. No attractive interactions are introduced.

The movement of the bacteria has been modeled by Brownian dynamic (BD) simulation \cite{LOW94}. In these simulations, the trajectories of the particles are obtained by integrating the Langevin equation. Thus, the trajectory of the center of mass and orientation of its longitudinal axis of an individual bacterium $i$, defined by the vectors $\textbf{r}_i$ and $\textbf{u}_i$, evolves in the time according to the following set of equations:

\begin{equation}
\label{eq2b}
\begin{split}
{\bf r}_i^{||}(t+\Delta t) =
{\bf r}_i^{||}(t)+\frac{D_{i||}}{k_BT} {\bf F}_i^{||}(t)\Delta t + \\
\,\,\,\,\,\,\,\,\,\, + (2D_{i||} \Delta t)^{1/2} R^{||} {\bf \hat{u}}_i(t)
\end{split}
\end{equation}

\begin{equation}
\begin{split}
{\bf r}_i^{\perp}(t+\Delta t) =    {\bf r}_i^{\perp}(t)+
\frac{D_{i\perp}}{k_BT} {\bf F}_i^{\perp}(t)\Delta
t+ \\
\,\,\,+ (2D_{i\perp} \Delta t)^{1/2} R^{\perp} \textbf{\^{v}}_i(t) \\
\end{split}
\end{equation}

\begin{equation}
\begin{split}
\textbf{\^{u}}_i(t+\Delta t) = \textbf{\^{u}}_i(t)+
\frac{D_{i\vartheta}}{k_BT} {\bf T}_i(t)\times
\textbf{\^{u}}_i(t)\Delta t+ \\
\,\,\,+ (2D_{i\vartheta} \Delta t)^{1/2} R^{\vartheta}
\textbf{\^{v}}_i(t)
\end{split}
\end{equation}

\noindent being $\textbf{r}_i^{\parallel}$ and $\textbf{r}_i^{\perp}$ the projections of $\textbf{r}_i$ on the directions parallel and perpendicular to $\textbf{\^{u}}_i$, respectively. $\textbf{F}_i^{\parallel}$ and $\textbf{F}_i^{\perp}$ are the parallel and perpendicular components of the total force acting on $i$ and ${\bf T}_i$ is the total torque due to the interactions with other particles of the fluid \cite{VEG90}. The particle (cell) Brownian dynamics is induced through a set of independent gaussian random numbers of variance 1 and zero mean: $R^{\parallel}$, $R^{\perp}$ and $R^{\vartheta}$. $\textbf{\^{v}}_{i}$ is an unitary vector perpendicular to ${\textbf{\^{u}}}_i$. 

The diffusion coefficients, $D_{i\parallel}$, $D_{i\perp}$ and $D_{i\vartheta}$ were calculated by a method similar to that proposed by Bonet Avalaos et al \cite{BON94}. They were provided to us by Fabi\'an A. Garc\'ia Daza by private communication. These diffusion coefficients depend on the size of the particles and they must be calculated for each bacterium at each time step. The explicit expressions for the calculation of these diffusion coefficients for a given aspect ratio are:

\begin{eqnarray}
D_{\parallel}/D_0 &=&-0.0198\cdot ln(L^*)+ 0.0777+\frac{0.0437}{L^*}\nonumber\\
&&-\frac{0.0158}{L^{*2}}\nonumber\\
D_{\perp}/D_0 &=& -0.0119\cdot ln(L^*) + 0.0452 + \frac{0.0796}{L^*}\nonumber\\
&&-\frac{0.0190}{L^{*2}}\\
D_{\vartheta}\,\sigma^2/D_0 &=& -0.0002\cdot ln(L^*)+0.0012 - \frac{0.0243}{L^*}\nonumber\\
&&+\frac{0.3233}{L^{*2}}
+\frac{0.2597}{L^{*3}}-\frac{0.0483}{L^{*4}}\nonumber
\end{eqnarray}

\noindent depending on the diffusional parameter $D_0=D_ 0^*\sigma^2/\tau$, with $\tau$ the time unit. These diffusion coefficients are not the same that were employed in \cite{ACE18}, the ones used were proposed by Shimizu \cite{SHI62} for prolate spheroids. As we are going to show later, no qualitative differences were found. In all the simulations discussed here the time step was fixed to $\Delta t=10^{-3}\tau$.

\begin{figure}[!t]
	\center
	\includegraphics[width =\columnwidth]{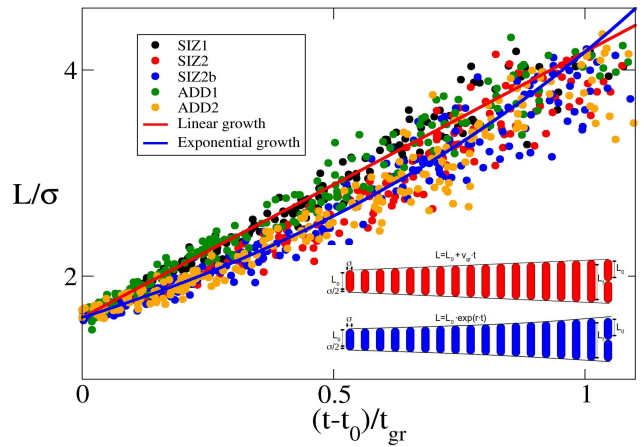}
 	\caption{Dependence of the elongation of individual bacterium with the time interval since the last division division ($t-t_0$) for scenarios $SIZ1$ (black circles), $SIZ2$ (red circles), $SIZ2b$ (blue circles), $ADD1$ (green circles) and $ADD2$ (orange circles). In all the cases $D^*_0=0.1$. The data are taken from random bacteria in colonies from 1 to 150 cells. Red and blue lines are the evolution of the elongation of an average bacterium in the case of linear and exponential growth, respectively. The inset shows the evolution of elongation of a single bacterium in the case of linear (red cells) and exponential (blue cells) growth, respectively. The main geometric characteristics of the cells are also indicated.}
	\label{fig1}
	\end{figure}

\begin{table*}[t]\label{tab1}
	\small
	\caption{Summary of the main characteristics of the scenarios explored in this paper. The acronyms for each scenario, the time dependence of the bacterial length and the magnitudes that in each scenario are affected by some level of stochastic dispersion are indicated. See the main text for a detailed description.}
	\vspace{-0.3cm}
	\begin{center}
		\begin{tabular*}{0.9\textwidth}{@{\extracolsep{\fill}}|c|c|c|c|}
			\hline \hline
			Scenario&Mechanism&Growth of individual bacteria&Variables with Stochastic Dispersion\\ 
			\hline
			$SIZ1$&Sizer&Linear with time& $v_{gr}$\\
			&           & $L(t) =L_0 +v_{gr}\cdot(t-t_0)$ & \\
			\hline
			$SIZ2$&Sizer&Exponential with time& $r$ \\
			&           & $L(t)=L_0\cdot exp(r\cdot(t-t_0))$ & \\
			\hline
			$SIZ2b$&Sizer&Exponential with time& $r$ and $L_f$\\
			\hline
			$ADD1$&Adder&Linear with time&$v_{gr}$ and $\Delta L$\\
			\hline
			$ADD2$&Adder&Exponential with time&$r$ and $\Delta L$\\
			
			\hline \hline
		\end{tabular*}		
	\end{center}
\end{table*}

An important component of our model is the modeling of bacterial elongation and division cycles. In our previous work \cite{ACE18} all the particles grew by polar lengthening at constant velocity $v_{gr}$ from a fixed initial elongation $L_0$. When the particles reached an aspect ratio $L^*_f =2L^*_0=2(L_0/\sigma+1)$ they divided into two identical particles, each with an initial aspect ratio $L^*_0$, and the same orientation as the parent particle (see Fig. 1 of \cite{ACE18} for more details). As mentioned previously, the purpose of this work is to examine how different lengthening or division scenarios, which have been proposed in the literature as a possible homeostatic mechanism to maintain stable the size distribution in a bacterial population, affect some collective properties of the microcolony. To do this, we have focused on scenarios analogous to those classically denoted as sizer and adder.

Firstly, we have carried out simulations where the elongation velocity of each particle $v^m_{gr}$ is selected at random at the moment of the division from a Gaussian distribution with mean $v_{gr}$ and relative standard deviation $s/v_{gr}=0.1$. In this scenario the elongation of the cylindrical part of each bacteria grows linearly with the time, $L(t) =L_0 +v^m_{gr}\cdot(t-t_0)$, being $L_0=L(t_0)$ the elongation of the bacteria just after a division, that is the same for all the particles. $t_0$ is the instant when the last division occurred. When a bacterium reaches an aspect ratio $L^*_f =2L^*_0=2(L_0/\sigma+1)$ it is divided into two identical cells as in \cite{ACE18}. This sizer scenario has been denoted as $SIZ1$.

We have defined another additional sizer scenario, named $SIZ2 $. In this scenario the elongation of the cylindrical part of an individual bacterium grows exponentially: $L(t)=L_0\cdot exp(r^m\cdot(t-t_0))$, being $r^m$ the elongation rate of bacterium $m$. $r^m$ is chosen at random from a Gaussian distribution centred in $r$ and relative standard deviation $s/r=0.1$. In this scenario, the division also occurs when $L^*_f =2L^*_0$ as in $SIZ1$. A comparison in the growth of an individual bacterium with linear and exponential elongation is shown in the inset of Fig,\ref{fig1}. This figure also indicates the geometric characteristics of the cells. 

As variation of the previous, in scenario $SIZ2b$ the final elongation of each bacterium $L_{f,m}$ is chosen from a Gaussian distribution centered in $L_f=2\cdot L_0+\sigma$ and relative standard deviation $s/L_f=0.1$. As each bacterium divides into two identical bacteria of elongation $L_{0,d}=0.5\cdot(L_{f,m}-\sigma)$, in this scenario the initial elongation is not the same for all bacteria. Subscripts $m$ and $d$ indicate mother and daughter bacterial cell, respectively.

We continue by describing the adder-type scenarios used in this work. In the first ($ADD1$), each individual bacterium divides when the elongation of its cylindrical part is increased a quantity $\Delta L_m$ from its initial value $L_{0,m}$. $\Delta L_m$ is chosen from a Gaussian distribution centred in $\Delta L=L_0+\sigma$ and relative standard deviation $s/\Delta L=0.1$. Again, the result of the division are two identical bacteria with elongation $L_{0,d}=0.5\cdot(L_{f,m}-\sigma)$, being in this case 
$ L_{f,m} = L_{0,m}+\Delta L_m$. As in $SIZ1$, in this scenario bacteria show linear lengthening, being $v_{gr}$ chosen from the same Gaussian distribution. Finally, $ADD2$ differs from $ADD1$ which now the elongation of each bacterium depends exponentially on time, like in $SIZ2$, $L_m(t)=L_{0,m}\cdot exp(r^m·(t-t_0))$. 
Table \ref{tab1} summarizes the main characteristics of the described scenarios.

Figure \ref{fig1} shows the increase in cell elongation through a reproductive cycle for the different scenarios described above. This figure shows the elongation as a function of the time elapsed since the last division, taken from bacteria at different times of the microcolony development. For reference, the average bacterial elongation is also shown as a function of time since the last division in the cases of linear and exponential growth. In this figure is possible to observe the differences between linear (scenarios $SIZ1$ and $ADD1$ ) and exponential ($SIZ2$, $SIZ2b$ and $ADD2$) growth, as well as the effect of the dispersion in $L_f$ (scenarios $SIZ2b$, $ADD1$ and $ADD2$). But probably it is more relevant that, here, it is verified that the stochastic dispersion of $v_ {gr}$, $r$ and $L_f$ causes that, although the average behavior is discernible between different scenarios, all sizes can be observed in all scenarios at a given interval from the start of the simulation. This is more evident in the case of microcolonies with many cells. This will be relevant to understanding the result and conclusions of our work, as we will see later. 

As mentioned, the main objective of this study is to explore the influence of the different scenarios described above on the structure and morphology of microcolonies. For this, we have calculated a set of observables, averaging typically over 80 runs in each case. Therefore, we have estimated de amount of biomass in the microcolony, $bms(t)$ as

\begin{equation}\label{eq2}
bms(t)= \sum_{i=1}^{N(t)} L^*_i(t)
\end{equation}

\noindent being $N(t)$ the number of cells at time $t$. As $N(t)$ and the aspect ratio of each particle $L^*_i(t)$  vary over time, $bms(t)$ also depend on time.

To determine the shape of the microcolony, we have calculated the ellipsoid that best fits the distribution of particles. For this we have determine the components of inertia tensor as $I_{\alpha,\beta}=1/N(t) \sum_{i=1}^{N(t)} \left(\delta_{\alpha,\beta} (\sum_{k=\alpha,\beta} r^k_{i})-  r^{\alpha}_{i} r^{\beta}_{i}\right)$. Here the $\alpha$ and $\beta$ indicate the coordinates $x$ or $y$, $\delta_{\alpha,\beta}$ is the Kronecker delta and $r^{\alpha}_i$ is the corresponding coordinate of the vector from the center of mass of the microcolony to the position of the bacterium $i$. Diagonalizing this tensor is possible to calculate the two semi-axes, $a>b$, of the ellipse that best fit the distribution of bacteria in the microcolony \cite{KAR07}. With them, it is possible to define an eccentricity parameter to measure how the shape of the microcolony differs from a circle:

\begin{equation}\label{eq3}
ecc^2(t) = 1-\frac{b^2}{a^2}
\end{equation}

With this definition $ecc^2(t)$ tends to $0$ for circular microcolonies.

As a global measure of the compactness of the microcolony, we have calculated the density as $\rho(t)=bms(t)/A_e(t)$, with $A_e(t)$ the area of the ellipse resulting from the diagonalization of the inertia tensor described above. To characterize the orientational correlation of the cells, we have calculated the nematic order parameter $S_2(t)$. This nematic order parameter is obtained with the standard procedure of diagonalizing a symmetric tensor traceless build with the orientation vectors of all the particles. For the particular case of two-dimensional systems, the expression for this tensor reads \cite{ALL93,MER92}

\begin{equation}\label{eq3b}
{\bf{Q}}=\frac{1}{N(t)}\left<\sum_{i=1}^{N(t)}(2\hat{u}_{i}(t)\hat{u}_{i}(t)-\bf{I})\right>,
\end{equation}

These fourth parameters, $bms(t)$, $ecc^2(t)$, $\rho(t)$ and $S_2(t)$, are time-dependent global indicators, changing over the development of the microcolony. In addition, we have calculated another set of observables that provide information about the internal structure of the microcolony at given instants, in contrast to the global information obtained from those previously defined. Thus, at given biomass values, we have determined the coverage profile $g(r_{cm})$. As it has been reported previously \cite{ACE18,LOB21}, relevant information about the internal structure of the microcolony could be obtained from this function. $g(r_{cm})$ is defined as the fraction of the surface covered by bacteria at a distance $r_{cm}$ from the microcolony center of mass. To calculate this function, we have generated a high number of random points at a distance $r+dr$ from the microcolony center of mass, evaluating $g(r_{cm})$ as the fraction of these points that fall into the area occupied by a bacterium. 

Finally, we have calculated the orientational distribution function between two particles. For two-dimensional systems as the interest here, this correlation function is defined as $g_2(r)=\left\langle (2(\textbf{\^{u}}_i\cdot\textbf{\^{u}}_j)-1)\delta(r_{ij}-r)\right\rangle$, with $r_{ij}$ the distance between the particles $i$ and $j$, $\delta(·)$ the Dirac delta, and the angular brackets meaning average over pair of particles and different trajectories. This function provides information about the distance dependence of the averaged relative orientation between the particles, allowing evaluation of the size of possible nematic domains.

\section{Results}\label{results}

We have applied the different scenarios described above to situations previously addressed in our previous studies \cite{ACE18,LOB21}. Therefore, we have tried to model bacteria with similar characteristics to \textit{Pseudomona putida}. For this Gram-negative bacterium, an aspect ratio of approximately $2.6$ has been determined experimentally \cite {ROD16}. Consequently, the particle elongation and aspect ratio of $L_0 = 1.6\sigma$ and $L^*_0 = 2.6 $ have been set as a reference values. As condition initial, all the simulations start with a single bacterium of aspect ratio $L^*_0$. According to \cite{ACE18}, colony morphology and structure are highly dependent on the relationship between the bacterial diffusion, elongation, and division times. We summarized this relationship by defining the parameter $\Gamma$:
 
\begin{equation}\label{eq6}
\Gamma =\frac{t_{dif}}{t_{gr}}
\end{equation}

\noindent being $t_{dif}$ the average time required by an isolated particle of constant aspect ratio $L^*_0$ to diffuse a distance $\sigma$ by brownian diffusion, and $t_{gr}$ the time need by an average bacterium to reach the aspect ratio $L^*_f$ from its initial aspect ratio. Both in $SIZ1$ and $ADD1$, $t_{gr}=(L_0+\sigma)/v_{gr}$, while for $SIZ2$, $SIZ2b$ and $ADD2$ $t_{gr}=1/r\cdot ln((2L_0+\sigma)/L_0)$ . $\Gamma$ is depending both on the diffusional parameter $D_0$ and on $v_{gr}$ or $r$, for linear or exponential growth respectively. Table \ref{tab2} shows the values of $D^*_0$, $v_{gr}$ and $r$, as well as the resulting values of $\Gamma$ for the cases considered in this study, corresponding to the different scenarios detailed in the previous section.

\begin{table}[t]\label{tab2}
		\small
	\caption{Values of $D^*_0$, $v_{gr}$ (in units of $\sigma/\tau$), $r$ (in units of $\tau^{-1}$) and $\Gamma$ in the various cases studied in this paper. Numbers in brackets represent alternative values used in scenarios $SIZ1$ and $ADD1$ for the indicated value of $\Gamma$.}
		\begin{tabular}{|c|c|c|c|}
		\hline 
		\multicolumn{1}{|c|}{Scenario} & $D^*_0$ & $v_{gr}\cdot\tau/\sigma$ & $\Gamma$\tabularnewline
		\hline 
		\multirow{5}{*}{$SIZ1$,$ADD1$} & \multirow{5}{*}{0.1(0.5)} & 0.0007(0.00352) & 0.01\tabularnewline
		\cline{3-4} 
		&  & {0.007(0.0355)} & 0.1\tabularnewline
		\cline{3-4} 
		&  & 0.07(0.353) & 1\tabularnewline
		\cline{3-4} 
		&  & 0.35(1.755) & 5\tabularnewline
		\cline{3-4} 
		&  & 1.05(5.27) & 15\tabularnewline
		\hline 
		\multicolumn{1}{|c|}{} &  & $r\cdot\tau$ & \tabularnewline\hline
		\multirow{5}{*}{$SIZ2$, $SIZ2b$, $ADD2$} & \multirow{5}{*}{0.1} & 0.000263 & 0.01\tabularnewline
		\cline{3-4} 
		&  & 0.0027 & 0.1\tabularnewline
		\cline{3-4} 
		&  & 0.028 & 1\tabularnewline
		\cline{3-4} 
		&  & 0.13 & 5\tabularnewline
		\cline{3-4} 
		&  & 0.39 & 15\tabularnewline
		\hline 
	\end{tabular}
\end{table}

\begin{figure}[!t]
	\center
	\includegraphics[width =\columnwidth]{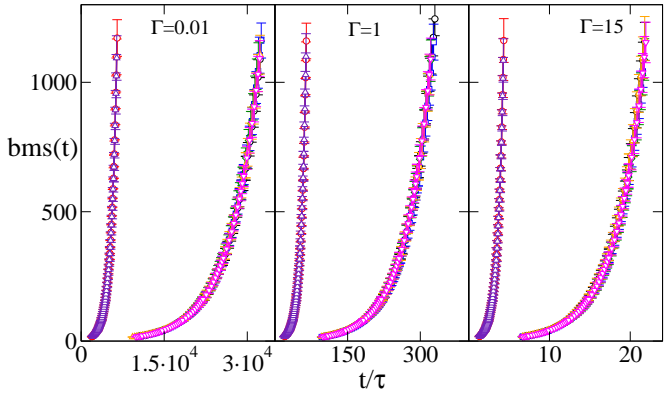}
	\caption{Biomass of the microcolony $bms(t)$  as a function of the time $t/\tau$ for values of $\Gamma= 0.01$ (left), 1 (center) and 15 (right). Each panel display results obtained in scenarios $SIZ1$ with $D^*_0=0.1$ (black line and circles), $SIZ1$ with $D^*_0=0.5$ (red line and circles), $ADD1$ with $D^*_0=0.1$ (orange line and triangles up), $ADD1$ with $D^*_0=0.5$ (violet line and triangles up), $SIZ2$ (blue line and squares), $SIZ2b$ (green line and diamonds) and $ADD2$ (magenta line and triangles down). In these last three cases $D^*_0=0.1$.}
	\label{fig2}
\end{figure}

Figure \ref{fig2} shows the dependence of biomass with time throughout the evolution of the microcolony for all the scenarios indicated in table \ref{tab1} and for $\Gamma=0.01, 1$ and $15$. As a first result, it is interesting to verify that, regardless of whether the elongation of each bacterium is linear or exponential, the growth of the biomass of the full colony follows the expected exponential law. It is also relevant that in all the cases where $D^*_0$ and $\Gamma$, and therefore $t_{gr}$, coincide, regardless of the scenario by which the bacteria grow and divide, the evolution of biomass over time collapses into a single curve. Cases with the same $\Gamma$ but different $D^*_0$ (and therefore different $t_{gr}$) do not show the same evolution of $bms(t)$, consequence of that in each case the exponential growth rate is $ln(2)/t_{gr}$.

\begin{figure}[!t]
	\center
	\includegraphics[width =\columnwidth]{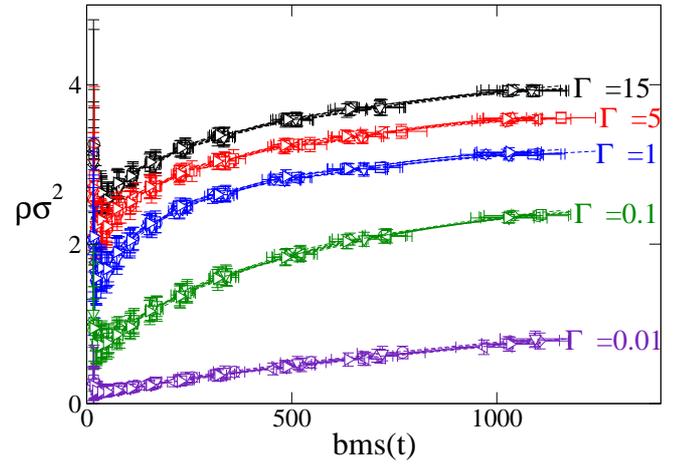}
	\caption{Density of the microcolony in reduced units ($\rho\sigma^2$) as a function of the biomass $bms(t)$. Black, red, blue, green and violet symbols refer to $\Gamma=$ 15, 5, 1, 0.1 and 0.01, respectively. Circles and squares are for scenario $SIZ1$ with $D^*_0=0.1$ and $0.5$, respectively. Triangles left and triangles down are for scenario $ADD1$ with $D^*_0=0.1$ and $0.5$, respectively. Diamonds, triangles up and triangles right are for scenarios $SIZ2$, $SIZ2b$ and $ADD2$, respectively. In these last three cases $D^*_0=0.1$.}
		\label{fig3}
\end{figure}

\begin{figure*}[t]
	\center
	\includegraphics[width =2\columnwidth]{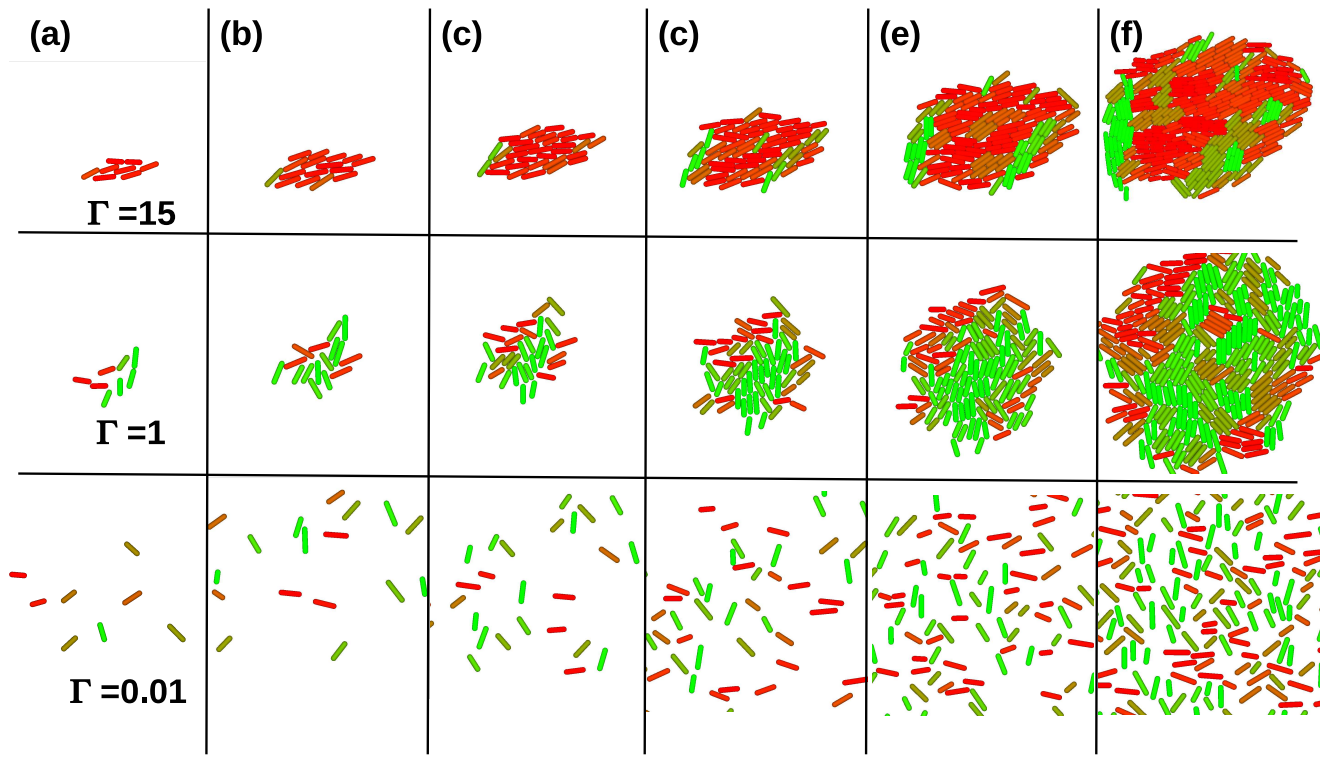}
	\caption{Snapshots of microcolonies with approximate biomass and number of bacteria $(bms,N_p)$ (a) $(22,8)$, (b) $(56,16)$, (c) $(95,32)$, (d) $(187,64)$, (e) $(374,127)$ and (f) $(899,316)$. From top to bottom we show the evolution in the case of $\Gamma= 15, 1$ and $0.01$, respectively. All these snapshots have been obtained with scenario $SIZ1$ and $D^*_0=0.1$. Particle color indicates orientation in a scale ranging from green (vertical) to red (horizontal).}
	\label{figconfis}
\end{figure*}

\begin{figure}[!t]
	\center
	\includegraphics[width =\columnwidth]{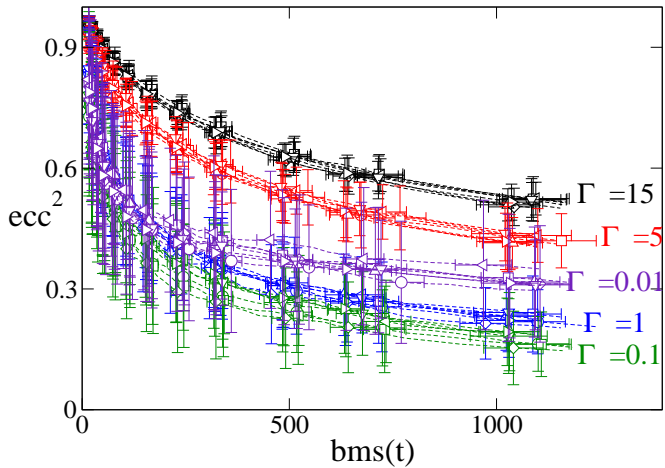}
	\caption{Eccentricity of the microcolony ($ecc^2(t)$) as a function of the biomass $bms(t)$. Symbols are the same than in figure \ref{fig3}.}
	\label{fig4}
\end{figure}

\begin{figure}[!t]
	\center
	\includegraphics[width =\columnwidth]{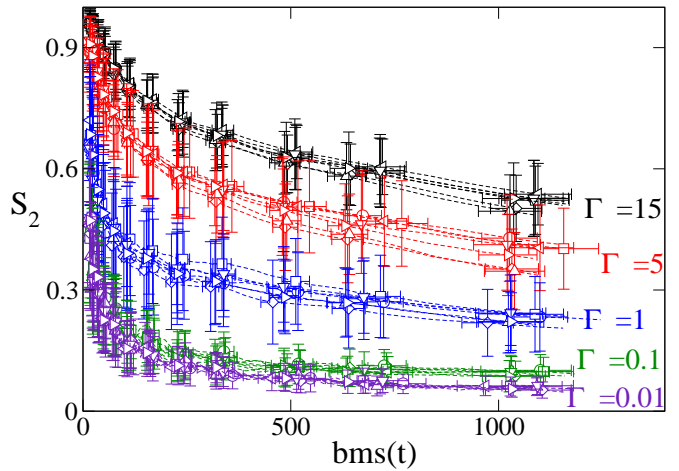}
	\caption{Nematic order parameter the microcolony ($S_2(t)$) as a function of the biomass $bms(t)$. Symbols are the same than in figure \ref{fig3}.}
	\label{fig5}
\end{figure}

\begin{figure}[t]
	\center
	\includegraphics[width =\columnwidth]{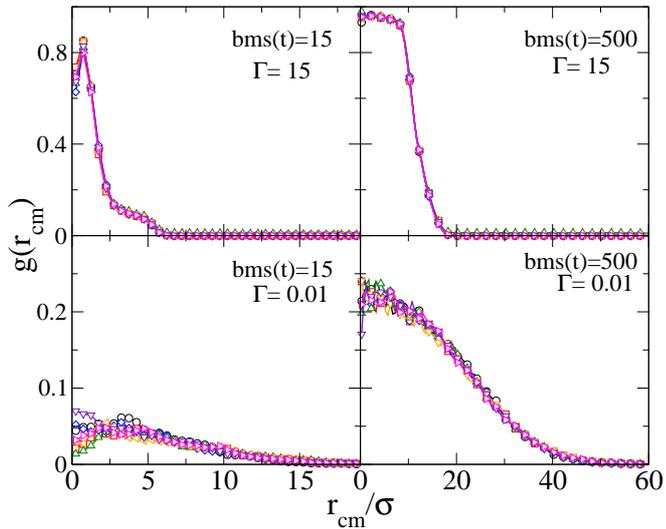}
	\caption{Surface coverage profiles $g(r_{cm})$ for microcolonies with $bms(t)=15$ and $500$ (left and right column, respectively) and $\Gamma=15$ and $0.01$ (top and
		and bottom row, respectively). Each panel displays results obtained in scenarios $SIZ1$ with $D^*_0=0.1$ (black line and circles), $SIZ1$ with $D^*_0=0.5$ (red line and squares), $ADD1$ with $D^*_0=0.1$ (orange line and triangles left), $ADD1$ with $D^*_0=0.5$ (violet line and triangles down), $SIZ2$ (blue line and diamonds), $SIZ2b$ (green line and triangles up) and $ADD2$ (magenta line and triangles right). In these last three cases $D^*_0=0.1$.}
	\label{fig6}
\end{figure}

\begin{figure}[!t]
	\center
	\includegraphics[width =\columnwidth]{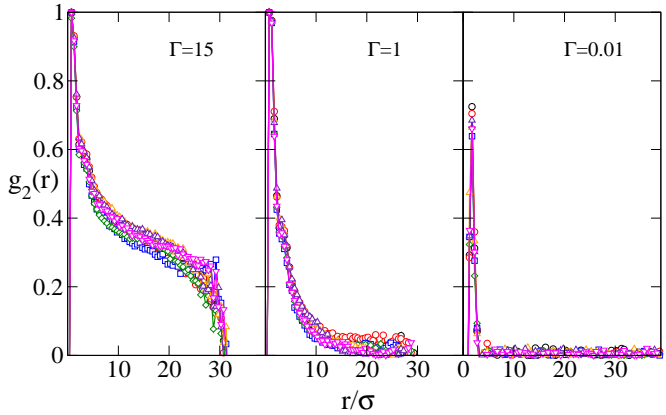}
	\caption{Orientational correlation function $g_2(r)$ for microcolonies with $bms(t)=500$ and $\Gamma=0.01, 1$ and $15$ (left, middle and right panel, respectively). Each panel displays results obtained in scenarios $SIZ1$ with $D^*_0=0.1$ (black line and circles), $SIZ1$ with $D^*_0=0.5$ (red line and squares), $ADD1$ with $D^*_0=0.1$ (orange line and triangles left), $ADD1$ with $D^*_0=0.5$ (violet line and triangles down), $SIZ2$ (blue line and diamonds), $SIZ2b$ (green line and triangles up) and $ADD2$ (magenta line and triangles right). In these last three cases $D^*_0=0.1$.}
	\label{fig7}
\end{figure}

It was stated in \cite{ACE18} that the structure and morphological properties of the simulated microcolonies only depended on $\Gamma$, regardless of the actual value of $D_0$, but no systematic proof was provided. We present here these evidences. Thus, figures \ref{fig3} to \ref{fig7} show that, for a given value of biomass in the microcolony, the structural and morphological properties calculated in this work are independent of the value of $D^*_0$, once a value of $\Gamma$ is set. Indeed, for simulations with $SIZ1$ and $ADD1$ scenarios, $\rho$ (Fig.\,\ref{fig3}), $ecc^2$ (Fig.\,\ref{fig4}) and $S_2$ (Fig.\,\ref{fig5}) at a given values of the biomass,  are independent of the value of $D^*_0$ for same value of $\Gamma$. Hence, for these three magnitudes, the simulation results obtained in the framework of these scenarios using $D^*_0 = 0.1$ and $0.5$ but keeping constant the value of $\Gamma$ (15, 5, 1, 0.1 or 0.01) are practically indistinguishable. A similar coincidence is observed when the comparison is done with structural properties. Thus, figures \ref{fig6} to  and \ref{fig7}, is observed that, for scenarios $SIZ1$ and $ADD1$ and a given value of $\Gamma$, the results obtained by simulation for $g(r_m)$ and $g_2(r)$ collapse in a single curve, regardless the value of $D^*_0$ and $t_{gr}$.

Thereby, the discussion in the previous paragraph supports the idea that, for a given scenario, the morphological and structural properties are just a function of $\Gamma$, beyond the values taken separately for $D^*_0$ and $t_{gr}$, as previously proposed in \cite{ACE18}. But even, the observation of the figures \ref{fig3} to \ref{fig7} suggests that the collective properties of the colony do not depend on the reproduction mechanism of the individual bacterium, being controlled only by the value of $\Gamma$. This is verified by the coincidence of the different observables for given values of $bms$ and $\Gamma$, regardless of the scenario in which bacterial growth and division are simulated. To analyze this coincidence in more detail, and to discuss the general characteristics of early biofilm growth, we will now detail the behavior of the different observables, highlighting the collapse of the studied cases into a single curve for given values of $\Gamma$ and $bms$.

Figure \ref{fig3} shows how density evolves with biomass. Density is a measure of how compact a growing colony is, with lower values indicating that the colony is more spread out. In this figure, it is appreciated how the greater the value of $\Gamma $, the greater the value $\rho $ for a given
 biomass. This is a consequence of the fact that in cases with low values of $\Gamma$, particle diffusion dominates over bacterial growth, favoring the dispersion of bacteria on the surface, as indicated in \cite{ACE18}. In any case, for all values of $\Gamma$, the density grows monotonically with $bms(t)$.  At high values of $\Gamma$ we explain this by the interaction between the growing bacteria, pushing each other. In contrast, at low values of $\Gamma$ this increase in the density with the biomass is a consequence of the filling of the inner of the microcolony by cell reproduction. 
 
In Fig.\,\ref{figconfis} is presented the growing sequence for microcolonies with $\Gamma= 15$, $1$ and $0.01$, respectively. In the three cases in scenario $SIZ1$ with $D^*_0=0.1$. These snapshot sequences show how, for $\Gamma=15$, the microcolony maintains a compact configuration from a very low number of cells, and throughout the sequence. This compactness is reduced for $\Gamma=1$, where in any case, the colony is still observed as a cluster of cells along the whole sequence. The situation changes radically for $\Gamma=0.01$, where the cells at the first stages disperse over the surface, increasing the local density in the last configurations shown due to the effect of continuous cell reproduction. This phenomenology is consistent with the behavior presented for density. 

In addition, it can be seen from Fig.\,\ref{fig3} that, for a given value of $\Gamma$, the evolution of $\rho(t)$ with $bms(t)$ is independent of the homeostatic mechanism, the type of growth, or the value of $D^*_0$. Moreover, once $\Gamma$ is fixed and independent of scenarios or $D^*_0$ values, the dependence of density on biomass collapses into a single line, at least within the range of microcolony sizes studied.

Figure \,\ref{fig4} shows the dependence of the eccentricity parameter of the ellipsoid that best fits the particle distribution, $ecc^2(t)$, for all the cases discussed in this work. As can be seen, for all $\Gamma$, $ecc^2$ decreases when $bms(t)$ increases. This implies an evolution from initial elliptical aggregates toward more circular microcolonies. This tendency is more pronounced for small values of $\Gamma$. In contrast, at high $\Gamma$, the microcolony maintains an appreciably non-circular shape, with $ecc^2(t)$ values above 0.5 for $\Gamma=15$. This is also confirmed in the configurations shown in figure \ref{figconfis}. It is noteworthy that the dependence of $ecc^2 (t) $ with $\Gamma$ is not monotonous, being greater for $\Gamma=0.01$ than for $\Gamma=1$, and $0.1$, especially for large values of $bms(t)$. This is a consequence of the more irregular shape of the particle distributions in the latter case, due to the dispersion of particles. As before, the dependence of $ecc^2(t)$ with $bms(t)$ is independent of the scenario or value of $D^*_0$ with which they were obtained, once the value of $\Gamma$ has been set.

We are going to discuss the orientation correlation between the particles. Thus, figure \ref{fig5} shows the dependence of the nematic order parameter with the biomass for different values of $\Gamma$, in all the scenarios explored in this work. This nematic order parameter, widely used in the study of liquid crystals, provides information about the collective orientation of the particles. It takes values close to the unity if the particles are preferentially oriented in a given direction, and zero if the particles are oriented at random. In this figure it is observed as while $S_2(t)$ remains high for the larger values of $\Gamma$, for $\Gamma\leq 1$  $S_2(t)$ decreases very fast when $bms(t)$ grows. This means, as it is reflected in the configurations of Fig\,\ref{figconfis}, that for high $\Gamma$ the microcolony shows a relevant level of orientational correlation, which is completely lost for lower values of $\Gamma$. An intermediate situation is observed for $\Gamma=1$. In this case for high values of $bms(t)$ the level of global orientational order is low ($S_2(t)\sim0.3$), but in Fig.\,\ref{figconfis} it is possible to observe the existence of small nematic domains, with a high level of local orientational order. We will come back to this issue later.

In this figure, for $S_2(t)$, it is again confirmed that for a given value of $bms(t)$ the results only depend on $\Gamma$. Thus, independently of the homeostatic mechanism chosen, whether the lengthening is linear or exponential with time, or the net value of $D^*_0$, the results collapse into a single curve for each value of $\Gamma$. There are small differences, which in principle could be attributed to the statistical error, which in some cases is high due to the small number of particles reached by our simulations. The similarities between the dependence of $ecc^2$ and $S_2$ on $bms(t)$ are also very remarkable. This similarity is such that for high values of $\Gamma$ the curves are almost coincident, although for low values of $\Gamma$ the differences are significant. This could indicate the existence of some kind of universal law relating $ecc^2$ and $S_2$, and probably other quantities, such as the density shown in figure \ref{fig3}. Confirmation of this possible universal relationship would require further theoretical and simulation studies.

So far, we have discussed magnitudes that give global information on the evolution of the microcolony throughout its growth. We will now focus on another set of observables that provide information about the internal structure and organization of the early biofilm at selected times. We start by discussing the behaviour of the surface coverage profile $g(r_{cm})$. For this discussion we have selected two situations: one with low biomass ($bms(t)=15$), corresponding to approximately $4$ bacteria, and another situation with biomass $bms(t)=500$ and approximately $128$ bacteria. These two situations are representative of stages with different characteristics in the evolution of the microcolony. 

$g(r_ {cm})$ indicates the fraction of points at a given distance from the centre of mass of the microcolony $r_{cm}$ that are covered by a bacterium.  This function measures the level of cell scattering on the surface, being another indicator of the compactness of the microcolony. As illustrated in Figure \ref{fig6}, surface coverage $g(r_{cm})$ strongly depends on the value of $\Gamma$. Thus, this figure shows that for $\Gamma = 15$ (top row), at low value of $bms(t)$ the central area of the microcolony is practically covered by bacteria, with values of $g(r_ {cm})$ very close higher than $0.8$. From this core, $g(r_ {cm})$ drops sharply to the edge of the microcolony. At a later stage ($bms(t)=$ 500), the situation is qualitatively the same, with a heavily covered central core and a sharp drop in coverage at the edge of the colony.

In contrast, for $\Gamma = 0.01$ we observe a completely different behavior. In this case (bottom row of figure \ref{fig6}), at a low value of $\Gamma$ the coverage in the center of the colony is very low (below $0.1$), with a slight drop over long distances. At the highest value of $bms(t)$ the coverage in the central part of the microcolony increases, and a smooth decay is still observed up to long distances. This behavior, previously reported in \cite{ACE18}, is a reflex of the behavior described from the visual inspection of the configurations shown in figure \ref{figconfis}. Thus, at high values of $\Gamma$, where cell elongation is dominant over diffusion, the microcolony grows as a compact and crowded aggregate. In contrast, for low values of $\Gamma$ diffusion dominates over elongation and cells in an early stage spread out over the surface. At a later stage, the cell's reproduction cycle fills the inner part of the whole. These two regimes were referred to in \cite{ACE18} as closed and open growth, respectively. Remarkably, the change between these two regimes is very abrupt. For the values of $\Gamma$ discussed in this work, the open growth regime has been observed only for $\Gamma=0.01$. This is coherent with the strong differences in the density of the microcolony obtained in simulations with $\Gamma=0.1$ and $\Gamma=0.01$ observed in Fig.\,\ref{fig3}.

Finally, in figure \ref{fig7} the orientational correlation function $g_2(r)$ is compared for cases with $\Gamma =15$, $1$ and $0.01 $. This function reports on the orientation correlation of particles that are at a certain distance $ r $, providing information on the size of possible nematic domains. We only discuss situations with large biomass values ($bms(t) = 500$), when the microcolony is large enough that the collective properties are already consolidated. In this figure it is observed how, for $\Gamma = 15$ in the left panel of Fig.\,\ref{fig7}, the strong orientational correlation between the particles in contact ($g_2(r=\sigma)=1$) decays very slowly with $r$. In this case, $g_2(r)$ maintains significant values, with a slow decay until a sharp drop at the edge of the microcolony. This indicates the existence of nematic domains with dimensions in the order of the size of microcolonies. These nematic domains, which can be observed in Fig.\,\ref{figconfis}, are formed by bacteria with similar orientations. For $\Gamma=1$, middle panel of Fig.\,\ref{fig7}, we observe a qualitatively similar situation. The orientational correlation at contact is very high, again $g_2(r=\sigma)=1$ indicates that bacteria at contact are parallel. From here, $g_2(r)$ decays with the interparticle distance, up to values close to 0 for distances in the order of ten bacterial diameters. In any case, at distances shorter than $5\sigma$ the orientational correlation is high enough to indicate the existence of nematic domains, with a number of bacteria of about ten. These nematic domains are significantly smaller than for $\Gamma=15$. Indeed, in Fig.\,\ref{figconfis} it is observed as for $\Gamma=1$ the microcolony is a set of many small nematic domains, while for $\Gamma=15$ the microcolony is made up of a few large nematic domains.

Similar to $g(r_{cm})$, for $\Gamma=0.01$ $g_2(r)$ shows a completely different behavior. At short distances, a peak in $g_2(r)$ indicates that the closer bacteria trend to be parallel. But this peak is not now strictly at contact (it appears at $r\sim 2\sigma)$, and it reaches a value lower than one. Hence, bacteria closer to each other can now have different orientations. At greater distances, the orientational correlation disappears. As a sign of an almost complete lack of orientation order, $g_2(r)$ dropped sharply to 0. This means that in this case there is no long-range orientational order, as was also reflected in the discussion of the nematic order parameter (Fig,\,\ref{fig5}), and as can be seen in Fig.\,\ref{figconfis}.
 
 As with other magnitudes described above, the results obtained for $g(r_{cm})$ and $g_2(r)$ are independent of the reproduction scenario studied. All the cases discussed in figures \ref{fig6} and \ref{fig7} collapse to a single curve, one time the values of $\Gamma$ and $bms(t)$ are fixed. The small differences observed in the different figures can be attributed to statistical error, due to the small number of bacteria reached in our simulations. 

\section{Discussion and Final Remarks}\label{Discussion}

In this study, we have confirmed some of the results that have been previously reported on the characteristics of biofilm development in its earliest stages, when they are still bidimensional microcolonies. Thus, modeling the bacteria with an aspect ratio equivalent to that of \textit{Pseudomonas putida}, we have found that, if the elongation of the cells is faster than diffusion, compact microcolonies are formed from the first moment. These microcolonies are ellipsoidal and show a high level of internal orientational correlation. In \cite{ACE18} this regimen was referred to as close growth. In contrast, if diffusion is dominant over cell elongation and division, the computer simulation results indicate that cells spread along the surface. At a later stage, the cells aggregate in a loose swarm as cell reproduction fills the inner regions of the bacteria distribution. In this case, the bacterial aggregates are less compact, with a tendency to present a circular shape and without internal orientational correlation. This mechanism has been called open growth \cite{ACE18}. 

In this work, we have verified that these different behaviours can be summarized with the $\Gamma$ parameter, previously introduced in \cite{ACE18}. Thus, the open growth regime appears in simulations with a very low $\Gamma$ parameter (of the order of $10^{-2}$). Note that the results only depend on the value of the $\Gamma$ parameter. In this study, we have verified that simulations with different rates of bacterial reproduction and diffusion coefficient, but the equal value of $\Gamma$, lead to a qualitatively identical situation, with matching values for the observables defined and calculated in this study.

The main objective of our study is to explore whether significant differences were found when cell division is modeled following a size or adder mechanism, and also to check the effect on the structural and geometric properties of the microcolony if bacteria lengthening is linear or exponential with time. Knowing which are the homeostatic mechanisms involved in keeping cell size stable over time is a hot topic on which recent publications have appeared with divergent conclusions \cite{FAC17,SAU16,TAH15,FAC19,XIA20}. We conclude that both the homeostatic mechanism and the time dependence of bacterial length change do not play a fundamental role in the structural and geometric properties of bacterial microcolonies. In particular, the homeostatic mechanism at work does not seem to influence the transition between closed and open growth regimes, which is apparently only controlled by the $\Gamma$ parameter. We understand that the main reason for this is the stochastic dispersion introduced in some of the bacterial properties in our simulation model. This stochastic variability is realistic, and has been found in experimental systems \cite{FAN19,FAC19} due to variation in environmental conditions, as well as the diversity of bacterial communities. This statistical dispersion diminishes the relevance of the differences introduced by linear or exponential lengthening, or by the homeostatic mechanism on the emergent properties of the bacterial communities studied in this work. 

Our results should not be interpreted in the sense of downplaying the problem of which is the dominant homeostatic mechanism (sizer, adder, timer, or a combination of some of them) in a given bacterial species to maintain stable size in the cells of a given population. This is a fundamental aspect to understand the biology of bacterial microcolonies, and more research will be necessary in the future to gain a better understanding of this issue. The conclusion that can be extracted from our work is that, although the homeostatic mechanism that is acting in each case has a great impact on the statistical characterization of the size distribution, and is a very interesting aspect of the individual cell biology, it does not look to be very relevant to explain collective properties of bacterial communities. At least, what concerns to structural and geometrical properties.

Our study may also be relevant for the development and improvement of future computer simulation models of bacterial communities. Our results limit the importance of factors such as homeostatic mechanisms, growth type or stochastic dispersion of cellular characteristics when simulating bacterial populations. We hope that this work will help future developments within this emerging branch of cell simulation.

\begin{acknowledgments}
	The authors acknowledge support from Consejer\'ia de Transformaci\'on Económica, Industria, Conocimiento y Universidades de la Junta de Andalucía/FEDER (project P20-00816), and from the Spanish Ministerio de Ciencia, Innovación y Universidades and FEDER (Project no. PGC2018-097151-B-I00). We are thankful to C3UPO for the HPC facilities provided.
\end{acknowledgments}

\nocite{*}
\bibliography{biblio}

\end{document}